\documentclass[aps,prc,twocolumn,groupedaddress,showpacs,showkeys]{revtex4-1}
\usepackage[english]{babel}
\usepackage{amssymb}
\usepackage{amsmath}
\usepackage{graphicx}

\begin{document}

\title{Particle-Hole Optical Model: Fantasy or Reality?}


\author{M.H. Urin}
\email[]{urin@theor.mephi.ru}
\affiliation{National Research Nuclear University ``MEPhI'', 31
Kashirskoye shosse, Moscow, 115409 Russia}


\date{\today}

\begin{abstract}
An attempt to formulate the optical model of particle-hole-type
excitations (including giant resonances) is undertaken. The model is
based on the Bethe--Goldstone equation for the particle-hole Green
function. This equation involves a specific energy-dependent
particle-hole interaction that is due to virtual excitation of
many-quasiparticle configurations and responsible for the spreading
effect. After energy averaging, this interaction involves an
imaginary part. The analogy between the single-quasiparticle and
particle-hole optical models is outlined.
\end{abstract}


\maketitle


Damping of giant resonances (GRs) is a long-standing problem for
theoretical studies. There are three main modes of GR relaxation:
(i) particle-hole (p--h) strength distribution (Landau damping),
which is a result of the shell structure of nuclei; (ii) coupling of
(p--h)-type states with the single-particle (s.p.) continuum, which
leads to direct nucleon decay of GRs; and, (iii) coupling of
(p--h)-type states with many-quasiparticle configurations, which
leads to the spreading effect. An interplay of these relaxation
modes takes place in the GR phenomenon.

A description of the giant-resonance strength function with exact
allowance for the Landau damping and s.p. continuum can be obtained
within the continuum-RPA (cRPA), provided the nuclear mean field and
p--h interaction are fixed \cite{ref1}. As for the spreading effect,
it is attempted to be described together with other GR relaxation
modes within microscopic and semimicroscopic approaches. The
coupling of the (p--h)-type states, which are the doorway states
(DWS) for the spreading effect, with a limited number of 2p--2h
configurations is explicitly taken into account within the
microscopic approaches (see, e.g., Refs.~\cite{ref2,ref3}). Some
questions to the basic points of these approaches could be brought
up: (i) ``Thermalization'' of the DWS, which form a given GR, i.e.
the DWS coupling with many-quasiparticle states (MQPS) (the latter
are complicated superpositions of 2p--2h, 3p--3h, \dots
configurations), is not taken into account. As a result, each DWS
may interact with others via 2p--2h configurations. Due to
complexity of MQPS one can reasonably expect that after energy
averaging the interaction of different DWS via MQPS would be close
to zero (the statistical assumption). (ii) The use of a limited
basis of 2p--2h configurations does not allow to describe correctly
the GR energy shift due to the spreading effect. The full basis of
these configurations should be formally used for this purpose. (iii)
With the single exception of Ref.~\cite{ref4}, there are no studies
of GR direct-decay properties within the microscopic approaches.

Within the so-called semimicroscopic approach, the spreading effect
is phenomenologically taken into account directly in the cRPA
equations in terms of the imaginary part of an effective s.p.
optical-model potential \cite{ref5,ref6}. Within this approach, the
afore-mentioned statistical assumption is supposed to be valid and
used in formulation of the approach. The GR energy shift due to the
spreading effect is evaluated by means of the proper dispersive
relationship, and therefore, the full basis of MQPS is formally
taken into account \cite{ref7}. The approach is applied to
description of direct-decay properties of various GRs (the
references are given in Ref.~\cite{ref6}). In accordance with the
``pole'' approximation used for description of the spreading effect
within the semimicroscopic approach, the latter is valid only in the
vicinity of the GR energy. However, for an analysis of some
phenomena it is necessary to describe the low- and/or high-energy
tails of various GRs. For instance, the asymmetry (relative to
90$^\circ$) of the ($\gamma \mathrm n$)--reaction differential cross
section at the energy of the isovector giant quadrupole resonance is
determined, in particular, by the high-energy tail of the isovector
giant dipole resonance \cite{ref8}. Another example is the
isospin-selfconsistent description of the IAR damping. In
particular, the IAR total width is determined by the low-energy tail
of the charge-exchange giant monopole resonance
\cite{ref9,ref9_5,ref10}.

In the present work, we attempt to formulate a model for
phenomenological description of the spreading effect on p--h
strength functions at arbitrary (but high enough) excitation
energies. The formulation of this semimicroscopic model (simply
called as the p--h optical model) is analogous to that of the
single-quasiparticle optical model \cite{ref11,ref12}. The
dispersive version of this model \cite{ref11} is widely used for
description of various properties of single-quasiparticle
excitations at relatively high energies (see, e.g.,
Ref.~\cite{ref13}). This model can be also used for description of
direct particle decay of subbarrier s.p. states \cite{ref14}.

The starting point in formulation of the single-quasiparticle
optical model is the Fourier-component of the Fermi-system
single-quasiparticle Green function, $G(x,x';\varepsilon)$, taken in
the coordinate representation (see, e.g., Refs.~\cite{ref11,ref12}).
By analogy with that, we start formulation of the p--h optical model
from the Fourier-component of the Fermi-system (generally, nonlocal)
p--h Green function, $\mathcal A(x,x';x_1,x'_1;\omega)$, also taken
in the coordinate representation. Being a kind of the Fermi-system
two-particle Green function (definitions see, e.g., in
Ref.~\cite{ref15}), $\mathcal A$ satisfies the following spectral
expansion:


\begin{widetext}
\begin{equation}
\mathcal{A}(x,x';x_1,x'_1;\omega) = \sum_s \left(
\frac{\rho_s^\ast(x',x)\rho_s(x_1,x'_1)}{\omega-\omega_s+i0}
-\frac{\rho_s^\ast(x'_1,x_1)\rho_s(x,x')}{\omega+\omega_s-i0}
\right). \label{eq1}
\end{equation}
\end{widetext}
Here, $\omega_s = E_s-E_0$ is the excitation energy of an exact
state $|s\rangle$ of the system and $\rho_s(x,x') = \langle s|
\widehat\Psi^+(x) \widehat\Psi(x')|0\rangle$ is the transition
matrix density ($\widehat\Psi^+(x)$ is the operator of particle
creation at the point $x$). In accordance with the expansion of
Eq.~(\ref{eq1}) the p--h Green function determines the strength
function $S_V(\omega)$ corresponding to an external (generally,
nonlocal) single-quasiparticle field $\widehat V = \int
\widehat\Psi^+(x) V(x,x') \widehat\Psi(x') dxdx' \equiv \left[
\widehat\Psi^+ V \widehat\Psi \right]$:
\begin{equation}
S_V(\omega) = -\frac{1}{\pi} \mathrm{Im} \left[ V^+ \mathcal
A(\omega) V \right], \label{eq2}
\end{equation}
where the brackets $\left[ \dots \right]$ mean the proper
integrations.

The free s.p. and p--h Green functions, $G_0(x,x';\varepsilon)$ and
$\mathcal A_0(x,x';x_1,x'_1;\omega)$, respectively, are determined
by the mean field (via the single-quasiparticle wave functions) and
the occupation numbers (only nuclei without nucleon pairing are
considered). Being determined by Eq.~(\ref{eq1}), the free
transition matrix densities $\rho_s^{(0)}(x,x')$ are orthogonal:
$\left[ \rho_s^{(0)\ast}\rho_{s'}^{(0)} \right]= \delta_{ss'}$. As
for the transition densities $\rho_s^{(0)}(x = x')$, which appear in
the spectral expansion for the free local p--h Green function
$A_0(x,x_1;\omega) = \mathcal A_0(x=x', x_1=x_1';\omega)$, this
statement is wrong. The RPA p--h Green function, $\mathcal
A_{RPA}(x,x';x_1,x'_1;\omega)$, is determined also by a p--h (local)
interaction $\mathcal F(x,x';x_1,x'_1) =
F(x,x_1)\delta(x-x')\delta(x_1-x'_1)$, which is responsible for
long-range correlations leading to formation of GRs. In particular,
the Landau--Migdal forces $F(x,x_1) \rightarrow F(x)\delta(x-x_1)$
are used in realizations of the semimicroscopic approach of
Refs.~\cite{ref5,ref6}. The RPA p--h Green function satisfies the
expansion, which is similar to that of Eq.~(\ref{eq1}). In such a
case, the RPA states $|d\rangle$ are the DWS for the spreading
effect. The local RPA p--h Green function $A_{RPA}(x,x_1;\omega)$
determined by the p--h interaction $F(x,x_1)$ is used for cRPA-based
description of the GR strength function corresponding to a local
external field $\widehat V = \int \widehat\Psi^+(x) V(x)
\widehat\Psi(x) dx$ \cite{ref1}.

The s.p. and p--h Green functions satisfy, respectively, the Dyson
and Bethe--Goldstone integral equations:
\begin{equation}
G(\varepsilon) = G_0(\varepsilon) + \left[ G_0(\varepsilon)
\Sigma(\varepsilon) G(\varepsilon) \right] \label{eq3}
\end{equation}
and
\begin{equation}
\mathcal A(\omega) = \mathcal A_{RPA}(\omega) + \left[ \mathcal
A_{RPA}(\omega) \mathcal P(\omega) \mathcal A(\omega) \right],
\label{eq4}
\end{equation}
where
\begin{equation}
\mathcal A_{RPA}(\omega)=\mathcal A_0(\omega)+\left[\mathcal
A_0(\omega)\mathcal F\mathcal A_{RPA}(\omega)\right]. \label{eq5}
\end{equation}
The self-energy operator $\Sigma(x,x';\varepsilon)$ and the specific
p-h interaction (polarization operator) $\mathcal
P(x,x';x_1,x'_1;\omega)$ describe the coupling, correspondingly, of
single-quasiparticle and (p--h)-type states with proper MQPS.
Analytical properties of $G$ and $\Sigma$ are nearly the same. A
similar statement can be made for $\mathcal A$ and $\mathcal P$. The
quantities $\Sigma(\varepsilon)$ and $\mathcal P(\omega)$ both
exhibit a sharp energy dependence due to a high density of poles
corresponding to virtual excitation of MQPS. Concluding
consideration of the basic relationships given above in a rather
schematic form, we present the alternative equation for the p--h
Green function:
\begin{equation}
\mathcal A(\omega) = \mathcal A_0(\omega) + \left[ \mathcal
A_0(\omega) \left(\mathcal F+\mathcal P(\omega)\right) \mathcal
A(\omega) \right], \label{eq6}
\end{equation}
which follows from Eqs.~(\ref{eq4}), (\ref{eq5}).

Since the density of MQPS, $\rho_m$, is large and described by
statistical formulae, only the quantities $\bar
\Sigma(x,x';\varepsilon)$ and $\bar {\mathcal
P}(x,x';x_1,x'_1;\omega)$ averaged over an interval $J \gg
\rho_m^{-1}$ can be reasonably parameterized. As applied to $\bar
\Sigma(\varepsilon) = \Sigma\left(\varepsilon +
iJSgn(\varepsilon-\mu)\right)$, it is done, e.g., in
Refs.~\cite{ref11, ref12}:
\begin{equation}
\bar \Sigma(x,x';\varepsilon) = Sgn(\varepsilon-\mu)
\left\{-iw(x;\varepsilon)+p(x;\varepsilon)\right\}\delta(x-x').
\label{eq7}
\end{equation}
Here, $\mu$ is the chemical potential and $w(x;\varepsilon)$ is the
imaginary part of a (local) optical-model potential. Assuming that
the radial dependencies of $p$ and $w$ are the same, i.e.
$w(x;\varepsilon) \rightarrow w(r)w(\varepsilon)$ and
$p(x;\varepsilon) \rightarrow w(r)p(\varepsilon)$, the intensity of
the real addition to the mean field, $p(\varepsilon)$, has been
expressed in terms of $w(\varepsilon)$ via the corresponding
dispersive relationship \cite{ref11}. It is noteworthy, that the
optical-model addition to the mean field can be taken as the local
one, i.e. $\bar \Sigma(x,x';\varepsilon) \sim \delta(x-x')$, in view
of a large momentum transfer (of order of the Fermi momentum) at the
``decay'' of single-quasiparticle states into MQPS. The energy
averaged single-quasiparticle Green function $\bar
G(x,x';\varepsilon)$ satisfies the Eq.~(\ref{eq3}), which involves
in such a case the quantity $\bar \Sigma$ of Eq.~(\ref{eq7}).
Actually, $\bar G$ is the Green function of the Schr\"{o}dinger
equation, which involves the addition to the mean field considerated
above.

The energy-averaged polarization operator can be parameterized
similarly to Eq.~(\ref{eq7}):
\begin{widetext}
\begin{equation}
\bar{\mathcal P}(x,x';x_1,x'_1;\omega) = \left\{-i\mathcal
W(x,x';\omega)+P(x,x';\omega)\right\}\delta(x-x_1)\delta(x'-x'_1).
\label{eq8}
\end{equation}
\end{widetext}
Assuming that the coordinate dependencies of the quantities $P$ and
$\mathcal W$ are the same, i.e. $\mathcal W(x,x';\omega) \rightarrow
\mathcal W(x,x')\mathcal W(\omega)$ and $P(x,x';\omega) \rightarrow
\mathcal W(x,x')P(\omega)$, we can express $P(\omega)$ in terms of
$\mathcal W(\omega)$ via the corresponding dispersive relationship.
The example of such a relationship is given in Ref.~\cite{ref7}. In
accordance with Eqs.~(\ref{eq4}), (\ref{eq6}), (\ref{eq8}) the
energy-averaged p--h Green function satisfies the equivalent
equations:
\begin{equation}
\bar{\mathcal A}(\omega) = \mathcal A_{RPA}(\omega) + \left[
\mathcal A_{RPA}(\omega) \bar{\mathcal P}(\omega) \bar{\mathcal
A}(\omega) \right], \label{eq9}
\end{equation}
\begin{equation}
\bar{\mathcal A}(\omega) = \mathcal A_0(\omega) + \left[ \mathcal
A_0(\omega) \left(\mathcal F+\bar{\mathcal P}(\omega)\right)
\bar{\mathcal A}(\omega) \right]. \label{eq10}
\end{equation}
Formally, Eqs.~(\ref{eq8})--(\ref{eq10}) are the basic equations of
the p--h optical model. In particular, the energy-averaged strength
function is determined by Eq.~(\ref{eq2}) with the substitution
$\mathcal A(\omega) \rightarrow \bar{\mathcal A} (\omega)$.

To realize the model in practice, a reasonable parametrization of
$\mathrm{Im} \bar{\mathcal P}$ should be done with taking the
statistical assumption into account. For this purpose, we consider
the quantity $\mathcal A_{RPA}$ within the discrete--RPA (dRPA) in
the ``pole'' approximation. In accordance with Eq.~(\ref{eq1}), we
have
\begin{equation}
\mathcal A_{RPA}(x,x';x_1,x'_1;\omega) \rightarrow \sum_d
\frac{\rho_d^\ast(x',x)\rho_d(x_1,x'_1)}{\omega-\omega_d+i0}.
\label{eq11}
\end{equation}
The statistical assumption $\left[ \rho_d^\ast \bar{\mathcal P}
\rho_{d'} \right] \sim \delta_{dd'}$ is fulfilled, provided that:
(i) the intensity $\mathcal W(x,x';\omega)$ is nearly constant
within the nuclear volume, i.e. $\mathcal W(x,x';\omega) \rightarrow
\mathcal W(\omega)$; and, (ii) the dRPA transition matrix densities
are orthogonal, i.e. $\left[ \rho_d^\ast \rho_{d'} \right] =
\delta_{dd'}$. Under these assumptions, the solution of
Eq.~(\ref{eq9}) can be easily obtained in the pole approximation:
$\bar{\mathcal A}(\omega) =  \mathcal A_{RPA}(\omega +i\mathcal
W(\omega) - P(\omega))$. As a result, the energy-averaged strength
function is the superimposition of the DWS resonances:
\begin{equation}
\bar S_V(\omega) = -\frac{1}{\pi}\mathrm{Im}\sum_d
\frac{\left|\left[V
\rho_d\right]\right|^2}{\omega-\omega_d+i\mathcal
W(\omega)-P(\omega)}. \label{eq12}
\end{equation}
The quantity $2\mathcal W$ can be considered as the mean DWS
spreading width $\langle \Gamma_d^\downarrow \rangle $, which might
be larger than the mean energy interval between neighboring DWS
resonances.

A few points are noteworthy in conclusion of the above-given
description of the p--h optical model. Within the model the
spreading effect on formation of (p--h)-type excitations is
described phenomenologically in terms of the specific (p--h)
interaction $\bar{\mathcal P}$. Because the interference between the
spreading of particles and holes is taken into account by this
interaction, the latter cannot be expressed via the
single-quasiparticle self-energy operator $\bar\Sigma$. Formally,
the p--h optical model is valid at arbitrary (but high enough)
excitation energy. The low limit is determined by the possibility of
using the statistical formulae to describe the MQPS density. Within
the semimicroscopic approach, the substitution like $\omega
\rightarrow \omega+i\mathcal W(\omega) - P(\omega)$ is used in the
cRPA equations to take the spreading effect phenomenologically into
account in the ``pole'' approximation together with the statistical
assumption \cite{ref6, ref7}. Thus, the parametrization of $\mathcal
W(\omega)$ can be taken in the form widely used for the intensity of
the imaginary part of the effective s.p. optical-model potential in
implementations of the semimicroscopic approach. Within the s.p.
optical model the statistical assumption for ``decay'' of different
single-quasiparticle states with the same angular momentum and
parity into MQPS seems to be valid. At high excitation energies
$|\varepsilon-\mu|$, when the empirical value of $w(\varepsilon)$ is
comparable with the energy interval between the afore-mentioned
single-quasiparticle states, the empirical radial dependence $w(r)$
becomes nearly constant within the nuclear volume (see, e.g., Refs.
[14]).

The p--h optical model can be simply realized in terms of the
energy-averaged local p--h Green function $\bar A(x,x_1;\omega)$ to
describe the strength function of a ``single-level'' GR, because in
such a case there is no need for the statistical assumption. Being
more simple, the equations like (\ref{eq4})--(\ref{eq6}),
(\ref{eq8}), (\ref{eq10}) are actually the straight-forward
extension of the corresponding cRPA equations. In practice, within
the cRPA it is more convenient to use the equation for the effective
field $\bar V(x,\omega)$, which corresponds to a local external
field $V(x)$ and is determined in accordance with the relationship:
$\left[ V\bar A(\omega) \right] = \left[ \bar V(\omega) A_0(\omega)
\right]$. The effective field determines the strength function:
\begin{equation}
\bar S_V(\omega) = -\frac{1}{\pi}\mathrm{Im}\left[ V_0 A_0(\omega)
\bar V(\omega) \right], \label{eq13}
\end{equation}
and satisfies the equation:
\begin{equation}
\bar V(\omega) = V + \left[ \left(F+\bar \Pi(\omega)\right)
A_0(\omega) \bar V(\omega) \right]. \label{eq14}
\end{equation}
The energy-averaged local polarization operator can be parameterized
similarly to Eq.~(\ref{eq8}):
\begin{equation}
\bar \Pi(x,x_1;\omega) =
C\left\{-iW(x;\omega)+P(x;\omega)\right\}\delta(x-x_1). \label{eq15}
\end{equation}
Here, $C=300$\,MeV\,fm$^3$ is the value often used in
parametrization of the Landau--Migdal forces; $W$ and $P$ are the
dimensionless quantities, which can be parameterized as follows:
$W(x;\omega) \rightarrow W(r) W(\omega)$ and $P(x;\omega)
\rightarrow W(r) P(\omega)$, where $P(\omega)$ is determined by
$W(\omega)$ via the corresponding dispersive relationship
\cite{ref7}.

\begin{figure}
\includegraphics[scale=0.8]{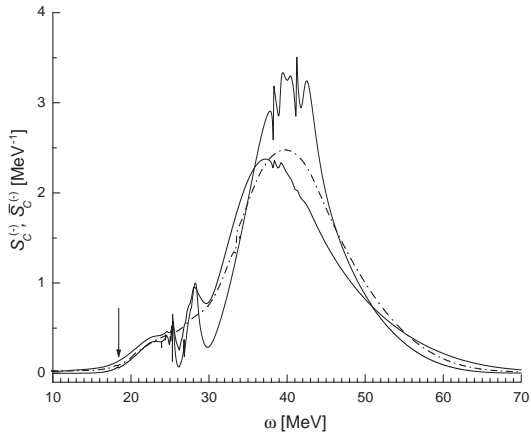}%
\caption{\label{fig1}The ``Coulomb'' strength function calculated
for the $^{208}$Pb parent nucleus within the cRPA (thin line), the
semimicroscopic approach (dash--dotted line), and p--h the optical
model (full line). The arrow indicates the IAR energy.}
\end{figure}

Due to strong coupling with s.p. continuum, the high-energy GRs
(they are mostly the overtones of corresponding low-energy GRs) can
be roughly considered as the ``one-level'' ones. Being the IAR
overtone, the charge-exchange (in the $\beta^{-}$-channel) giant
monopole resonance (GMR$^{(-)}$) is related to these GRs. Within the
isospin-selfconsistent description of the IAR damping
\cite{ref9_5,ref10}, the low-energy ``tail'' of the GMR$^{(-)}$ in
the energy dependence of the ``Coulomb'' strength function $\bar
S_C^{(-)}(\omega)$ determines the IAR total width $\Gamma_A$ via the
nonlinear equation:
\begin{equation}
\Gamma_A = 2\pi S_A^{-1}\bar S_C^{(-)}(\omega = \omega_A).
\label{eq16}
\end{equation}
Here, $S_A \simeq \left( N-Z \right)$ is the IAR Fermi strength,
$\omega_A$ is the IAR energy, and the ``Coulomb'' strength function
corresponds to the external field $V(x) \rightarrow V_C^{(-)} =
\left( U_C(r) - \omega_A +\frac{i}{2} \Gamma_A \right) \tau^{(-)}$,
where $U_C(r)$ is the mean Coulomb field. Strength function $\bar
S_C^{(-)}(\omega)$ exhibits a wide resonance corresponding to the
GMR$^{(-)}$. In Fig.~\ref{fig1}, we present the strength function
calculated for the $^{208}$Pb parent nucleus within: (i) the cRPA
(in such a case the strength function $S^{(-)}_C(\omega=\omega_A)$
determines the IAR total escape width found without taking the
isospin-forbidden spreading effect into account \cite{ref16}); (ii)
the semimicroscopic approach \cite{ref10} and, (iii) the p--h
optical model by Eqs.~(\ref{eq13})--(\ref{eq15}). All the model
parameters, parameterization of the imaginary part of the effective
single-quasiparticle optical-model potential $I(r;\omega)$
\cite{ref10} and parameterization of $W(r;\omega)$ in
Eq.~(\ref{eq15}) are taken the same in both approaches. The
intensities of $I(r;\omega)$ and $W(r;\omega)$ are chosen to
reproduce in calculations the observable total width of the
GMR$^{(-)}$ in $^{208}$Bi ($\simeq 15$ MeV). Both approaches lead to
the similar results, which are not exactly the same for the
low-energy ``tail'' of the GMR$^{(-)}$ at $\omega \simeq \omega_A$.
Irregularities in the energy dependence of $\bar S_C^{(-)}(\omega)$
calculated within the p--h optical model are explained by the fact
that the GMR$^{(-)}$ can be roughly considered as the ``one-level''
one.

%



In the present work, the optical model of particle-hole-type
excitations has been formulated in terms of the energy-averaged
nonlocal particle-hole Green function. The equation for this Green
function involves a specific energy-dependent particle-hole
interaction, which is due to virtual excitation of
many-quasiparticle configurations. The intensity of the imaginary
part of this interaction should be taken nearly constant within
nuclear volume to satisfy the statistical assumption on the
independent spreading of different particle-hole-type states which
form a given giant resonance. The strength function of the
``single-level'' giant resonance can be described in terms of the
energy-averaged local particle-hole Green function.

Along with numerical realizations, the particle-hole optical model
can be extended to describe direct particle decays of giant
resonances. These points are under consideration.

In conclusion, one can say that in formulation of the particle-hole
optical model we are on a way from fantasy to reality.

The author thanks M.L.~Gorelik for the calculation leading to the
results presented in Fig.~\ref{fig1}, and I.V.~Safonov for his kind
help in preparing the manuscript.

This work is partially supported by RFBR under grant
no.~09-02-00926-a.


%



\end{document}